# DISLOCATION AVALANCHE CORRELATIONS

Jérôme Weiss[a] and M.-Carmen Miguel[b]

[a] *Laboratoire de Glaciologie et Géophysique de l'Environnement, CNRS UMR 5183, 54 rue Molière, BP 96, 38402 St Martin d'Hères Cedex, France*
[b] *Departament de Fisica Fonamental, Factultat de Fisica, Universitat de Barcelona, Av. Diagonal 647, 08028 Barcelona, Spain*

*Abstract*
Recently, mechanical tests on ice as well as dislocation dynamics simulations have revealed that plastic flow displays a scale-free intermittent dynamics characterized by dislocation avalanches with a power law distribution of amplitudes. To further explore the complexity of dislocation dynamics during plastic flow, we present a statistical analysis of dislocation avalanche correlations and avalanche triggering. It is shown that the rate of avalanche triggering immediately after any avalanche is larger than the background activity due to uncorrelated events. This self-induced triggering increases in intensity, and remains over the background rate for longer times, as the amplitude of the mainshock increases. This analysis suggests that stress redistributions and the associated collective dislocation rearrangements may be responsible for aftershock triggering in the complex process of plastic deformation.

*1. Introduction*

Acoustic emission (AE) measurements performed during the creep of ice single crystals as well as dislocation dynamics simulations have shown that dislocations can move in a scale-free intermittent manner characterized by dislocation avalanches with power law distributions of amplitudes, $P(A) \sim A^{-\tau}$ [1]. This suggests a close-to-critical state for the dislocation ensemble during plastic deformation. Other deformation induced processes, such as fracture, display similar complexity. At geophysical scales, in addition to power law distributions of earthquake amplitudes, complexity of fracture and faulting is expressed by complex time patterning and interactions. Earthquakes trigger aftershocks with a rate that decays in time as a power law, $n(t) = K(t+c)^{-p}$, where $n(t)$ is the rate of aftershock triggering after a given mainshock, $p$ is a characteristic exponent close to 1, and $K$ and $c$ are constants (see e.g. [2]). Whereas the value of $p$ is independent of the amplitude of the mainshock, large earthquakes trigger many more aftershocks than smaller ones, a fairly intuitive result. Thus the constant $K$ is proposed to scale with the earthquake amplitude $A$ as $A^\alpha$ or, equivalently, after introducing the earthquake magnitude $M=\text{Log}(A)$, as $K \sim 10^{\alpha M}$, where $\alpha \sim 1$ [2, 3]. This slow and scale-free decay, which has been tentatively ascribed to mechanisms such as sub-critical cracking or fatigue [4], is rather unusual, since many physical systems relax instabilities through an exponential decay, $n(t)=K\exp(-t/t_0)$. At much smaller scales, aftershock triggering in the microfracturing of rocks has been documented by means of acoustic emission experiments [5, 6], though the data were not very conclusive about the most appropriate decay law (power law or exponential) in this case.

Previous work has revealed a time clustering of dislocation avalanches [7], as well as a complex space/time coupling that can be interpreted as the result of a cascade process where avalanches increase the occurrence probability of subsequent avalanches in their vicinity [8]. Here, we investigate in more detail these avalanche correlations by performing an analysis of triggered events (aftershocks) for two types of datasets: (i) acoustic emission measurements performed during the creep of ice single crystals, and (ii) model data obtained from 2D simulations of collective dislocation dynamics.

*2. Methods and results*

As detailed elsewhere [9], fast and collective motions of dislocations (dislocation avalanches) generate acoustic waves whose properties can be recorded by a piezoelectric transducer in an acoustic emission experimental setup. In particular, the occurrence time of an avalanche can be determined to a high resolution (0.1μs for the present work), as well as the amplitude of the acoustic wave $A$, which is proportional to the total area browsed by the dislocations during the avalanche. In our experiments, the magnitude $M = \text{Log}(A)$ of the avalanches ranges between –2.5 ($A \approx 3\times 10^{-3}$ Volts) and 1 ($A = 10$ Volts). Although avalanche locations can be estimated using an array of acoustic transducers [8], here we focus on dislocation avalanche correlations in the time domain.

Acoustic events are recorded during the creep deformation (constant load) of ice single crystals in a way detailed elsewhere [1, 7], and under conditions where inelastic deformation only occurs by dislocation glide. Plastic flow of ice single crystals is characterized by a very strong anisotropy: slip occurs essentially along basal planes of the hexagonal lattice [10], although non-basal loops of limited extension have also been observed [11]. Three different time constants are used by the recording equipment to automatically individualize events. These time constants are set first to include secondary echoes, due to waves reflected on the sample surface, within a single event. Consequently, the duration of an event has no clear relation to the duration of the avalanche. In addition, a "dead time" of 150 µs between the end of an event and the possibility of recording a new one is imposed. All of this precludes the analysis of a purely dynamic triggering of secondary avalanches by the stress wave generated by the mainshock, as the propagation time of an elastic wave (of velocity $v \approx 3900$ m/s) through the entire sample is about 20 µs. For normalized artificial acoustic sources (Nielsen test) of very large amplitude ($M$ close to 1), the average number of secondary echoes that were resolved as individual events by the AE system was 0.2 per mainshock. They occurred between 150 and 250 µs after the end of the mainshock.

From the AE catalogs corresponding to different creep tests, each of them containing between $10^4$ and $3\times10^5$ events, we calculated the average event rate $n_M(t)$ recorded after avalanches of magnitude $M$-$0.25 \leq M \leq M$+$0.25$, per mainshock and per unit time. The time origin ($t=0$) used for this calculation was the end of the mainshock of magnitude $M$. Consequently, $n_M(t)$ is independent of the distribution of event durations, at least for the first generation of aftershocks. Figure 1 shows three of the curves obtained for $M=0,-1,-2$. The triggering of secondary avalanches is clearly identified at small time scales, as $n_M(t)$ is larger than its background value due to uncorrelated events at longer times. This self-induced triggering decreases with time towards the background level. Note that the terms "mainshock" and "aftershock" or "secondary avalanches" are used here only with a statistical meaning: this analysis reveals dislocation avalanches correlations through time, but is unable to identify which event is triggering another one. In addition, an "aftershock" of the first generation can itself trigger aftershocks of the second generation, and so on and so forth. As for acoustic emission recorded during the microfracturing of rocks [5, 6], the present data do not allow to firmly conclude about the appropriate decay law for $n_M(t)$. This is related to a relatively low aftershock activity to background activity ratio for dislocation avalanches compared, for instance, to earthquakes. In the latter case, aftershock triggering following large earthquakes strongly dominates the background seismic activity over times much longer than the duration of the earthquake. In the present case, the impossibility of recording a new event earlier than 150 µs after the *end* of another event may modify the estimation of $n_M(t)$, especially for aftershocks of the second generation or above. As an example, the minima of $n_M(t)$ observed around 350 µs might result from the impossibility of recording any activity during 150 µs after the end of the "first generation" of aftershocks.

Nevertheless, Figure 1 clearly shows that avalanches of large magnitude trigger, on average, many more secondary events than smaller magnitude ones. This can be quantified by calculating the average number of aftershocks triggered by avalanches of magnitude $M$, $N_M$, from the integration of the part of $n_M(t)$ which is above the background activity. As observed for earthquakes [2, 3], $N_M$ scales as $10^{\alpha M}$, or $N_M \sim A^\alpha$ (see Figure 2), with $\alpha$ in the range 0.6-0.85 for the different tests. The values of $N_M$ observed for avalanches of large magnitude ($M=0.5\pm0.25$), i.e. $3.5 \leq N_M \leq 6$, should be compared to the average number of secondary echoes unfortunately individualized for the Nielsen tests of magnitude $M \approx 1$, i.e. 0.2 (see above). However, as these echoes occur within the time range 150-250 µs after the end of the mainshock (see above), they could marginally influence $n_M(t)$ in this time range ($0.2/10^{-4}$ s = 2000 event/s, to compare with $n_M(t) \approx$ 7500 event/s for $M=0.0\pm0.25$ within the time range 150 µs$<t<$250 µs; see Figure 1).

Further evidence for dislocation avalanche correlations and self-induced triggering can be obtained from the analysis of the output data of a dislocation dynamics model. This 2D numerical model, described in more detail elsewhere [12], accounts for the mutual long-ranged elastic interactions between dislocations, as well as for dislocation annihilation at short distances, and multiplication sources. The latter is phenomenologically taken into account through the random generation of new dislocation pairs with a rate proportional to the stress applied and to the system size. Aimed to simulate the plastic deformation of ice single crystals, this model represents a single slip system. The model degrees of freedom are dislocations (and not the atoms or molecules of the crystal) in the quasi-static approximation, and it is therefore unable to directly account for the acoustic waves generated by dislocation collective motions. Nevertheless, the model reasonably assumes that the acoustic wave amplitudes should be proportional to the collective velocity $V=\Sigma|v_i|$ of the fast moving dislocations, that is, the dislocations moving faster than if they were independently moving under the only action of the external stress $\sigma$. The velocity $V$ is calculated at each time step of the simulation in the corresponding time unit $t_0$ (see [1, 12] for details). Local maxima of recorded signal $V(t)$ (i.e. $V(t-1)<V(t)$ and $V(t+1)<V(t)$), expressing a reactivation of the collective dislocation dynamics, are here identified as dislocation avalanches of magnitude $M=\text{Log}(V)$. An aftershock analysis of the model data is performed in a similar manner to that described above for the experimental counterpart, except for the time origin ($t=0$) used in the calculation of $n_M(t)$ which now is simply the occurrence time of the event. An illustration of the results obtained is provided in Figure 3. These results share common properties with the results obtained from the experimental AE data. The average rate $n_M(t)$ is significantly larger than the background activity on small time scales, and decreases slowly towards this background value at longer times, as expected. Here we should emphasize, however, that the triggering of secondary avalanches is necessarily the result of the internal collective dislocation dynamics and cannot be influenced by the "experimental" procedure, or by other physical processes that are not implemented in the model. It is also worth noting a peculiar feature which shows up in Figure 3 and that was not observed on the AE data: whereas $n_M(t)$ increases with $M$ for $t>40t_0$, as for AE data, we observe that $n_M(t)$ decreases as a function of $M$ at shorter time scales (i.e. $t<30t_0$) for magnitudes larger than a threshold value $M_{th}$. This suggests an inhibition of the triggering activity immediately after very large avalanches responsible for the relaxation of high stress concentrations. Nevertheless, it might well be a spurious effect resulting as a consequence of the particular definition of an avalanche adopted in the model. Since the relaxation of the collective dislocation velocity immediately after a big burst is very slow (see the curve in the inset of Figure 3), it is very likely that the superposition of this slow relaxation with other events, possibly triggered right after the mainshock, is hiding the presence of their corresponding local maxima in the signal. Thus, by identifying avalanches as local maxima of the resulting signal, we may be underestimating the number of aftershocks. The threshold magnitude $M_{th}$ is system-size dependent since it increases from about 0.25 for a model of size 100$b$, with $b$ being the Burgers vector of a dislocation, to about 0.5 for a model of size 300$b$. The average number of aftershocks triggered by avalanches of magnitude $M$, $N_M$, is observed to increase with $M$ and fall within the range 3-5 for avalanches of large magnitude, in excellent agreement with the values observed for AE data.

*3. Discussion and conclusions*

As shown previously [1], plastic deformation is the result of a scale-free, intermittent, collective dislocation dynamics that suggests that the flow process may be happening in the vicinity of a critical state. The present analysis shows that this non-equilibrium dynamics is also characterized by the presence of dislocation avalanche correlations and self-induced avalanche triggering in the course of time. The enhancement of collective dislocation motions, and the corresponding acoustic activity, right after a previous avalanche has taken place is not merely the result of reflected elastic waves since: (i) the acoustic activity remains over the background level due to uncorrelated events for times as large as 0.01 s, i.e. about 1000 times larger than the travel time of the elastic waves through the sample, and (ii) a similar enhancement is observed in simulations of collective dislocation motion, unaffected by any experimental haphazard. Similar self-induced triggering is observed for earthquakes, where a similar relationship between the amplitude of the event and the average number of triggered aftershocks is observed, but with a much larger ratio between aftershock and background activity. Aftershock triggering is the result of stress redistributions following earthquakes. The persistence of aftershock activity over the years has been ascribed to mechanisms such as stress corrosion cracking or fatigue, that are not relevant in the study of dislocation dynamics. Similar stress redistributions follow dislocation avalanches. The induced dislocation rearrangements may occur initially through low-velocity dislocation motions (aseismic) which, after some delay, can trigger fast and cooperative re-activations of the dislocation dynamics, i.e. secondary avalanches. At −10°C, the velocity of an isolated, non-interacting, basal dislocation driven by a shear stress of 0.086 MPa (test shown on Figure 1) is about 1 µm/s [11]. The distance traveled by this dislocation in 0.01 s (the duration of the detectable aftershock activity) is 10 nm, i.e. 66$b$. This estimation, very small if compared to the distances of several mm over which space/time coupling of dislocation avalanches has been observed [8], suggests that the dislocation rearrangements taking place between avalanches occur at velocities intermediate between "seismic" velocities (a significant fraction of the acoustic wave velocity $v \approx$ 3900 m/s) and velocities of independent dislocations. The correlations discussed here provide a complementary illustration of the complexity of dislocation motion in the process of plastic deformation.

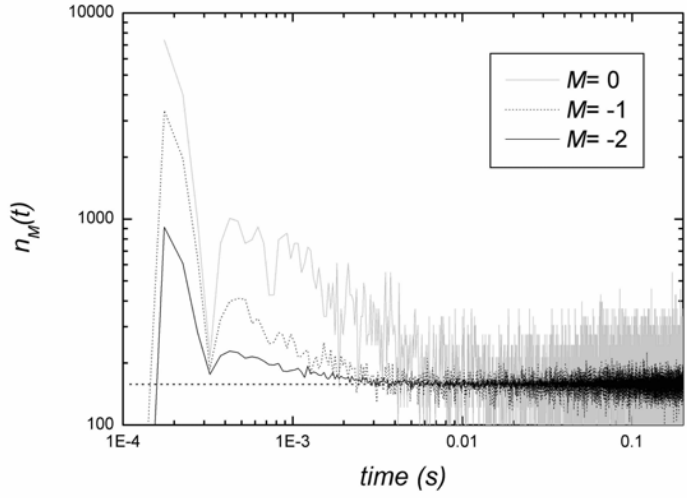

*Figure 1*. Average AE event rate $n_M(t)$ recorded after avalanches of magnitude $M\pm0.25$ per mainshock and per unit time (s), during a compression creep test of an ice single crystal (T=-10°C; resolved shear stress on the slip planes: 0.086 MPa; test duration: 3800 s; number of events detected: 281071). The background activity is represented by a horizontal dashed line.

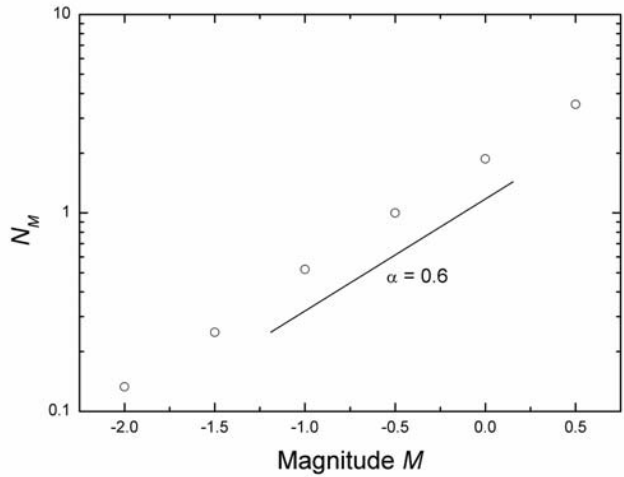

*Figure 2*. Average number of aftershocks triggered by dislocation avalanches of magnitude $M$, $N_M$, for the test described in Figure 1. $N_M$ increases with $M$ as $10^{\alpha M}$, i.e. as a power law of the avalanche amplitude $A$, $N_M \sim A^\alpha$, with $\alpha=0.6$.

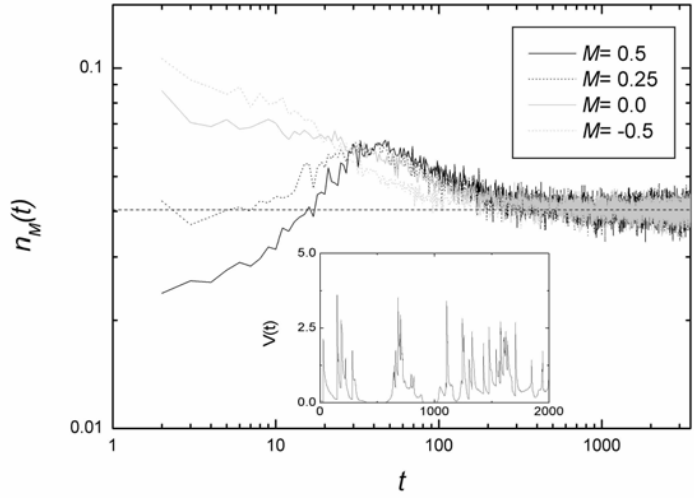

*Figure 3*. Average event rate $n_M(t)$ observed after avalanches of magnitude $M\pm0.25$, per mainshock and per unit time $t_0$, in a 2D dislocation dynamics simulation (model size: $100b$; shear stress: $0.01\sigma_0$, where $\sigma_0$ is the unit of stress used in the simulations; creation rate: $0.04t_0^{-1}$; duration of the simulation: $10^6$ $t_0$; number of events: 40413 ). The background activity is represented by a horizontal dashed line. In the inset we show a piece of the collective dislocation velocity $V(t)$ curve.